\documentclass[12pt,preprint]{emulateapj}

\usepackage{epsf}
\usepackage{color}
\usepackage{amsmath}
\usepackage{apjfonts}

\shorttitle{LYMAN ALPHA GALAXIES AT z $\sim$ 0.3}
\shortauthors{FINKELSTEIN ET AL.}

\newcommand{\sol}{$_{\odot}$}
\newcommand{\lya}{Ly$\alpha$}
\newcommand{\dust}{A$_{1200}$}
\def\arcs{\hbox{$^{\prime\prime}$}}
\def\arcm{\hbox{$^{\prime}$}}

\begin{document}
\slugcomment{Accepted for Publication in the Astrophysical Journal}

\title{EVOLUTION OF LYMAN ALPHA GALAXIES: STELLAR POPULATIONS AT z $\sim$ 0.3}

\author{\sc Steven   L.    Finkelstein\altaffilmark{1}}
\affil{George P. and Cynthia W. Mitchell Institute for Fundamental Physics and Astronomy, \\Department of Physics, Texas A\&M University, College Station, TX 77843}
\affil{Department  of  Physics,  Arizona  State University,  Tempe, AZ  85287} 

\author{Seth H. Cohen, Sangeeta Malhotra \& James E. Rhoads}   
\affil{School of Earth and Space Exploration,  Arizona  State University,  Tempe, AZ  85287}

\altaffiltext{1}{stevenf@physics.tamu.edu}

\begin{abstract}
We present the results of a stellar population analysis of 30 Lyman alpha emitting galaxies (LAEs) at z $\sim$ 0.3, previously discovered with the {\it Galaxy Evolution Explorer} ({\it GALEX}). With a few exceptions, we can accurately fit model spectral energy distributions to these objects, representing the first time this has been done for a large sample of LAEs at z $<$ 3, a gap of $\sim$ 8 Gyr in the history of the Universe.  From the 26/30 LAEs which we can fit, we find an age and stellar mass range of 200 Myr -- 10 Gyr and 10$^{9}$ -- 10$^{11}$ $M$\sol, respectively.  These objects thus appear to be significantly older and more massive than LAEs at high-redshift.  We also find that these LAEs show a mild trend towards higher metallicity than those at high redshift, as well as a tighter range of dust attenuation and interstellar medium geometry.  These results suggest that low-redshift LAEs have evolved significantly from those at high redshift.
\end{abstract}


\section{Introduction}

Over the past decade thousands of Lyman alpha emitting galaxies (LAEs) have been discovered from 2.4 $<$ z $<$ 6.96 (e.g., Cowie \& Hu 1998; Rhoads et al. 2000; Rhoads et al. 2001; Gawiser et al. 2006a; Kashikawa et al. 2006; Ouchi et al. 2008).  It was predicted that strong Ly$\alpha$ emission would be a signpost of primitive galaxies in formation, and thus these objects could represent some of the first galaxies in the Universe (Partridge \& Peebles 1967).  However, with the new results coming from stellar population modeling, the identification of LAEs as primitive galaxies has come into doubt.  Numerous studies have begun to detect dust in LAEs, which is an indicator that a prior generation of stars has lived and died, thus making these objects not primitive (Chary et al. 2005; Lai et al. 2007; Pirzkal et al. 2007; Nilsson et al. 2007; Kuiper et al. 2009; Yuma et al. 2009; Finkelstein et al. 2008, 2009).  While the presence of dust was originally proposed as the reason why LAEs were not detected 20 years ago, it is also possible that the dust can help enhance the value of the \lya~EW by selectively attenuating the continuum (Neufeld et al. 1991; Haiman \& Spaans 1999; Hansen \& Oh 2006).  In our previous work (Finkelstein et al. 2008, 2009), we demonstrated that dust enhancement from a clumpy interstellar medium (ISM) can help explain the spectral energy distributions (SEDs) of a sample of z $\sim$ 4.5 LAEs, with clumpy dust appearing to exist in 10 out of a sample of 14 LAEs, even though 12/14 of these LAEs were young ($<$ 15 Myr).  We also found plausible evidence for dust in every object in our sample, although their derived metallicities were rather low (with half of the sample having Z $<$ 0.2 $Z_{\odot}$).  This suggests that these objects are not primordial, as a previous generation of stars was likely responsible for the creation of the dust.

While we have estimates for these fundamental physical parameters of distant LAEs, it becomes interesting to study them at lower redshift where we can observe them in greater detail.  However, large samples of LAEs have only been studied down to z $\sim$ 3.1 (Gawiser et al. 2006a; McLinden et al. 2009; Nakamura et al. 2009), with a few recent samples currently under analysis at z $\sim$ 2 (Nilsson et al. 2008; Reddy et al. 2008).  Given the loss of CCD quantum efficiency at $\sim$ 3600 \AA~(and the atmospheric ultraviolet (UV) cut-off at $\sim$ 3100 \AA), detection of LAEs at z $\lesssim$ 2 is not possible from the ground.  However, LAEs could be detected using the space-based UV telescope, {\it Galaxy Evolution Explorer} ({\it GALEX}), at z $\lesssim$ 1.  Deharveng et al. (2008) recently published the discovery of 96 LAEs at 0.2 $<$ z $<$ 0.35 via {\it GALEX} spectroscopy, using data from the publicly available {\it GALEX} GR2 data release.  We present the results from a stellar population modeling analysis of 30 of these objects, with an emphasis on their differences from high-redshift LAEs.

Where applicable, we assume cosmology with $\Omega_{m}$ = 0.3, $\Omega_{\Lambda}$ = 0.7 and H$_{0}$ = 70 km s$^{-1}$ Mpc$^{-1}$ (c.f. Spergel et al. 2007).  All magnitudes in this paper are listed in AB magnitudes (Oke \& Gunn 1983).

\section{\lya~Galaxy Sample}

Deharveng et al. (2008; hereafter D08) recently published a {\it GALEX} catalog of low-redshift \lya~emitting galaxies, providing their positions, spectroscopic redshifts, \lya~line fluxes and EWs, and {\it GALEX} far-UV (FUV) and near-UV (NUV) broadband magnitudes.  To compare these objects to stellar population models, we require optical broadband fluxes.  We thus examined a subset of these LAEs in the Extended Groth Strip (EGS) and the Extended Chandra Deep Field -- South (ECDF--S), both of which have extensive multi-wavelength public datasets.
\begin{figure*}[th]
\epsscale{1.0}
\plotone{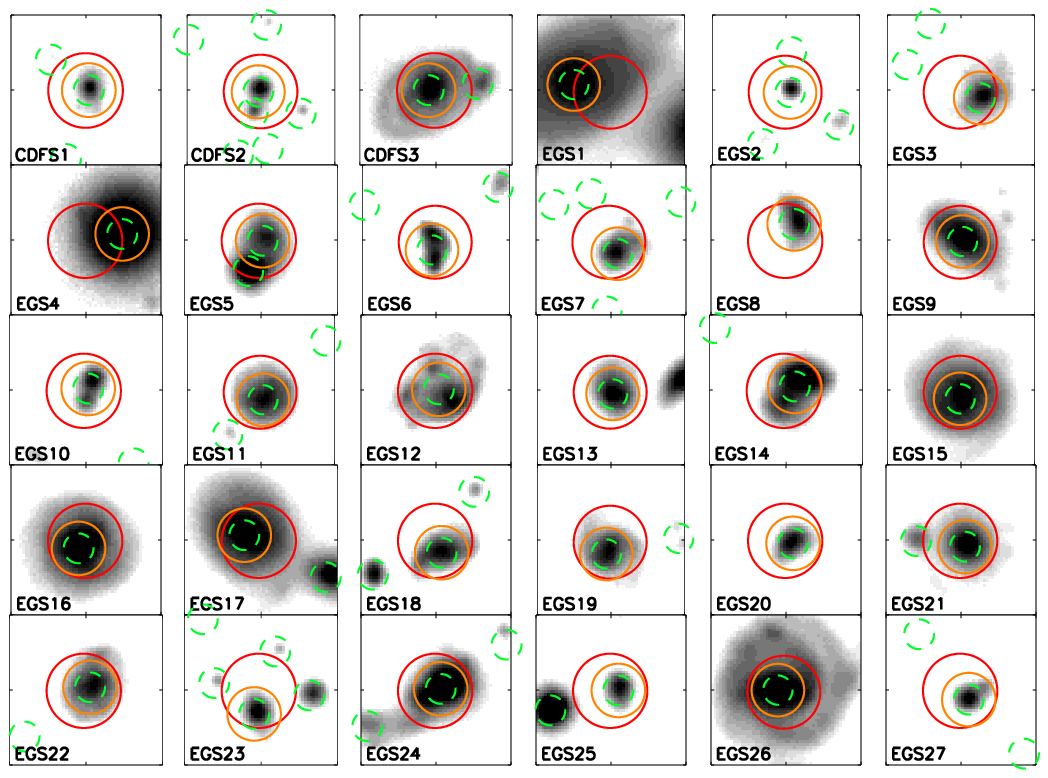}
\caption{Cutouts of the positions of the 30 LAE; MUSYC V-band for the CDFS, and CFHTLS i'-band for the EGS.  Large solid-line circles are centered on the {\it GALEX} positions of these objects as published in Deharveng et al. (2008), with radii of 2.5\arcs, comparable to the FWHM of the {\it GALEX} beam.  Small dashed-line circles denote all objects in the CFHTLS or MUSYC optical catalogs.  Smaller solid-line circles are centered on the object selected by our matching code to be the optical counterpart.  In the majority of cases, the circles all line up, meaning that the {\it GALEX} beam is centered on the optical counterpart.}\label{match}
\end{figure*}

The All-Wavelength Extended Groth Strip International Survey (AEGIS; Davis et al. 2007) covers the EGS from radio to X-ray wavelengths.  We used optical data from the Canada-France-Hawaii Telescope Legacy Survey\footnote[1]{Based on observations obtained with MegaPrime/MegaCam, a joint project of CFHT and CEA/DAPNIA, at the Canada-France-Hawaii Telescope (CFHT) which is operated by the National Research Council (NRC) of Canada, the Institut National des Science de l'Univers of the Centre National de la Recherche Scientifique (CNRS) of France, and the University of Hawaii. This work is based in part on data products produced at the Canadian Astronomy Data Centre as part of the Canada-France-Hawaii Telescope Legacy Survey, a collaborative project of NRC and CNRS.} (CFHTLS), which covers one deg$^{2}$ of sky (with seeing $\lesssim$ 1\arcs).  The 5 $\sigma$ limiting AB magnitudes of these data are:  u' (27.2), g' (27.5), r' (27.2), i' (27.0) and z' (26.0).  D08 publish the positions of 39 \lya~galaxies in the EGS, of which 30 fall in the CFHTLS area.  We then use positional matching to locate the optical counterparts to these 30 objects.  One object had no nearby optical counterpart, so we exclude it from further analysis, leaving us with 29 objects.  Given that {\it GALEX} has a spatial resolution of 4\arcs~and 5.6\arcs~FWHM in the FUV and NUV channels, respectively, {\it GALEX} positions can have high uncertainties relative to typical ground-based seeing ($\sim 5"$).  To account for this, we use a search radius  of 7\arcs, and then visually inspect the results to be sure that we are selecting the correct optical counterpart.  Figure 1 shows these results, displaying the CFHTLS i'-band image overplotted with large solid-line circles denoting the {\it GALEX} positions (with a 5\arcs~diameter), dashed-line circles denoting all objects from the CFHTLS i'-band selected catalog (2\arcs~diameter), and smaller solid-line circles encompassing the i'-band source that we selected as the optical counterpart.  In most cases, we see both large and small solid-line circles centered around an object. However, in two cases the {\it GALEX} position was between two or more sources, and in these latter cases we removed these objects from further consideration, leaving us with 27 EGS LAEs.  It is worth noting that some of the excluded objects could possibly be mergers, thus triggering star-formation and creating \lya~photons.  If this is the case, then the \lya~could be coming from one or both objects, however a detailed analysis of this possibility will await higher resolution FUV or \lya~imaging.
\begin{deluxetable*}{ccccccccccc}[h]
\tablecaption{Best-Fit Single Population Models of z $\sim$ 0.3 LAEs.}
\tablewidth{0pt}
\tablehead{
\colhead{IAU Format Name} & \colhead{Label} & \colhead{t$_{pop}$} & \colhead{Mass} & \colhead{Z} & \colhead{$\tau_{SFH}$} & \colhead{A$_{1200}$} & \colhead{q} & \colhead{EW} & \colhead{$\chi^{2}_{r}$} & \colhead{z$^{*}$}\\
\colhead{$ $} & \colhead{$ $} & \colhead{(Myr)} & \colhead{(10$^{9}$ M\sol)} & \colhead{(Z\sol)} & \colhead{(yr)} & \colhead{(mag)} & \colhead{$ $} & \colhead{(\AA)} & \colhead{$ $} & \colhead{$ $}
}
\startdata
J033211.92-280130.0&CDFS1&6000.0 $\pm$ 2695.5&\phantom{00}6.91 $\pm$ \phantom{00}0.47&0.005&4x10$^{9}$&0.20 $\pm$ 0.37&0.25 $\pm$ 1.38&155&36.11&0.22\\
J033256.66-275316.5&CDFS2&\phantom{00}45.0 $\pm$ \phantom{000}0.0&\phantom{00}4.56 $\pm$ \phantom{00}0.15&2.5&10$^{6}$&2.00 $\pm$ 0.00&0.00 $\pm$ 0.13&232&29.73&0.37\\
J033307.32-274432.6&CDFS3&3000.0 $\pm$ \phantom{0}250.0&107.70 $\pm$ \phantom{00}9.46&0.4&10$^{8}$&1.25 $\pm$ 0.35&2.00 $\pm$ 0.50&35&12.52&0.22\\
J142041.17+530650.2&EGS1&\phantom{0}453.5 $\pm$ 3273.3&\phantom{0}55.85 $\pm$ \phantom{0}88.35&1.0&10$^{7}$&4.50 $\pm$ 2.07&0.75 $\pm$ 1.38&240&16.76&0.20\\
J141743.43+522805.9&EGS2&\phantom{000}4.0 $\pm$ \phantom{000}0.0&\phantom{0}0.012 $\pm$ \phantom{0}0.002&0.2&10$^{5}$&0.40 $\pm$ 0.05&0.00 $\pm$ 1.50&221&371.17&0.21\\
J141814.11+522343.8&EGS3&\phantom{0}404.2 $\pm$ \phantom{0}719.9&\phantom{00}1.99 $\pm$ \phantom{00}0.67&0.4&10$^{7}$&1.75 $\pm$ 0.64&0.25 $\pm$ 0.75&311&9.50&0.21\\
J141710.64+531153.2&EGS4&5000.0 $\pm$ 1000.0&183.40 $\pm$ 131.80&0.2&10$^{8}$&0.80 $\pm$ 1.10&0.75 $\pm$ 0.25&147&16.34&0.21\\
J142226.54+523027.0&EGS5&2500.0 $\pm$ 2745.6&\phantom{00}3.91 $\pm$ \phantom{00}0.81&0.02&10$^{8}$&0.20 $\pm$ 0.64&0.25 $\pm$ 0.88&135&3.22&0.22\\
J141822.42+521823.6&EGS6&\phantom{0}321.0 $\pm$ \phantom{00}93.9&\phantom{00}1.11 $\pm$ \phantom{00}0.20&0.4&10$^{7}$&0.70 $\pm$ 0.37&0.00 $\pm$ 1.00&179&5.19&0.24\\
J141805.12+524507.1&EGS7&\phantom{0}286.1 $\pm$ \phantom{00}15.6&\phantom{00}1.40 $\pm$ \phantom{00}0.14&0.4&10$^{7}$&0.20 $\pm$ 0.23&1.50 $\pm$ 0.88&87&7.58&0.25\\
J141848.25+521756.1&EGS8&\phantom{0}360.2 $\pm$ \phantom{00}41.6&\phantom{00}1.28 $\pm$ \phantom{00}0.22&0.2&10$^{7}$&0.40 $\pm$ 0.37&0.25 $\pm$ 0.88&136&2.71&0.24\\
J141654.27+522440.0&EGS9&2500.0 $\pm$ 4625.0&\phantom{0}24.95 $\pm$ \phantom{0}18.19&0.2&10$^{8}$&1.00 $\pm$ 0.62&1.00 $\pm$ 0.63&115&2.98&0.25\\
J142043.29+524307.8&EGS10&\phantom{0}227.3 $\pm$ \phantom{0}206.6&\phantom{00}1.20 $\pm$ \phantom{00}0.20&0.4&10$^{7}$&0.60 $\pm$ 0.32&1.00 $\pm$ 1.38&96&1.08&0.25\\
J141914.72+522326.4&EGS11&\phantom{0}321.0 $\pm$ 1178.3&\phantom{00}3.47 $\pm$ \phantom{00}3.12&0.2&10$^{7}$&1.50 $\pm$ 1.01&1.00 $\pm$ 2.13&103&1.52&0.26\\
J142044.64+525006.5&EGS12&\phantom{0}286.1 $\pm$ \phantom{0}856.9&\phantom{00}3.47 $\pm$ \phantom{00}0.79&0.2&10$^{7}$&1.50 $\pm$ 0.62&1.00 $\pm$ 0.75&103&2.63&0.26\\
J141712.18+523556.6&EGS13&1800.0 $\pm$ 2106.9&\phantom{00}6.63 $\pm$ \phantom{00}1.21&0.02&10$^{8}$&0.60 $\pm$ 0.67&2.00 $\pm$ 1.00&69&3.46&0.26\\
J142124.54+523920.1&EGS14&3000.0 $\pm$ 1000.0&\phantom{0}37.39 $\pm$ \phantom{00}6.33&0.2&10$^{8}$&1.75 $\pm$ 1.06&1.00 $\pm$ 0.63&115&1.57&0.26\\
J141855.63+524927.7&EGS15&4000.0 $\pm$ 1500.0&\phantom{0}64.19 $\pm$ \phantom{0}10.83&0.4&10$^{8}$&0.30 $\pm$ 0.60&3.00 $\pm$ 1.75&62&5.40&0.27\\
J141952.94+522100.2&EGS16&3000.0 $\pm$ 4361.0&\phantom{0}62.34 $\pm$ \phantom{00}7.71&0.4&10$^{8}$&1.50 $\pm$ 1.20&1.50 $\pm$ 1.00&53&4.78&0.27\\
J141907.21+522108.3&EGS17&5000.0 $\pm$ 2179.7&151.77 $\pm$ \phantom{0}31.44&0.2&10$^{8}$&0.50 $\pm$ 0.97&1.50 $\pm$ 0.63&98&5.66&0.27\\
J141930.24+530204.8&EGS18&\phantom{0}286.1 $\pm$ \phantom{0}615.7&\phantom{00}2.22 $\pm$ \phantom{00}0.40&0.2&10$^{7}$&1.00 $\pm$ 0.51&1.50 $\pm$ 1.13&66&2.22&0.27\\
J141709.87+530511.3&EGS19&\phantom{0}360.2 $\pm$ 1339.5&\phantom{00}5.37 $\pm$\phantom{00} 2.97&0.02&10$^{7}$&1.75 $\pm$ 0.83&0.50 $\pm$ 1.25&225&4.43&0.27\\
J141936.54+530936.2&EGS20&\phantom{0}321.0 $\pm$ \phantom{00}37.0&\phantom{00}1.73 $\pm$ \phantom{00}0.27&0.02&10$^{7}$&0.60 $\pm$ 0.32&1.00 $\pm$ 1.38&103&2.11&0.27\\
J141937.64+523024.8&EGS21&1800.0 $\pm$ 4492.4&\phantom{0}14.72 $\pm$ \phantom{00}2.54&0.2&10$^{8}$&1.25 $\pm$ 0.67&1.50 $\pm$ 1.75&66&1.55&0.29\\
J141649.90+525019.9&EGS22&2500.0 $\pm$ 1856.9&\phantom{0}10.12 $\pm$ \phantom{00}3.52&0.02&10$^{8}$&0.20 $\pm$ 0.83&5.00 $\pm$ 2.13&58&2.35&0.29\\
J141855.91+525931.8&EGS23&\phantom{0}202.6 $\pm$ \phantom{0}615.7&\phantom{00}1.35 $\pm$ \phantom{00}0.32&0.2&10$^{7}$&1.00 $\pm$ 0.55&1.00 $\pm$ 0.63&104&2.72&0.29\\
J142208.10+525225.8&EGS24&\phantom{0}508.8 $\pm$ \phantom{00}27.7&\phantom{0}37.99 $\pm$ \phantom{00}6.58&0.2&10$^{7}$&2.50 $\pm$ 0.46&0.25 $\pm$ 0.13&566&6.27&0.31\\
J141616.62+530133.3&EGS25&\phantom{000}5.0 $\pm$ \phantom{0}317.8&\phantom{00}0.36 $\pm$ \phantom{00}1.14&1.0&10$^{5}$&3.50 $\pm$ 1.20&0.75 $\pm$ 0.63&123&72.04&0.33\\
J142154.50+522423.5&EGS26&10000. $\pm$ 5464.5&334.10 $\pm$ 129.40&0.005&4x10$^{9}$&1.75 $\pm$ 1.15&1.50 $\pm$ 0.50&63&4.64&0.35\\
J141717.29+523217.6&EGS27&\phantom{00}17.4 $\pm$ \phantom{00}59.9&\phantom{00}4.07 $\pm$ \phantom{00}2.72&0.005&10$^{5}$&4.00 $\pm$ 1.43&0.00 $\pm$ 0.13&400&228.35&0.46\\
\enddata
\tablecomments{$^{*}$The redshifts are GALEX-derived spectroscopic redshifts from Deharveng et al. (2008).  Coordinates and best-fit modeling results for our 30 LAEs.  The reported 1 $\sigma$ errors are discussed in \S 2.1.  In a few cases the error is formally zero, as these grids are not quite continuous, thus all the middle 68\% of the probability distribution all fell at the exact same age or dust value.}\label{fittab}
\end{deluxetable*}

We performed a similar procedure in the $\sim$ 30\arcm~$\times$ 30\arcm~ECDF-S, using data from the Multiwavelength Survey by Yale-Chile (MUSYC; Gawiser et al. 2006b), which cover the Extended ECDF-S to a 5 $\sigma$ depth of 26.0, 26.9, 26.4, 26.4, 24.6 and 23.6 AB mag in the U, B, V, R, I and z' filters, respectively.  Out of the 15 {\it GALEX} selected LAEs, we find that four are covered by the MUSYC data.  Two of these have unambiguous optical counterparts, while a third has one bright object at the center of the {\it GALEX} beam, with another faint object near the edge of the beam.  The fourth object has nothing at the center of the {\it GALEX} beam, with two objects at the edge.  We thus kept the first three objects, while excluding the fourth one.  Thus our final sample consists of 30 LAEs, 27 in the EGS and 3 in the ECDF-S.

\subsection{Stellar Population Modeling}
Our motivation for this study was to learn about the stellar population ages and masses of low-redshift LAEs.  To do this, we use the stellar population modeling code of Bruzual \& Charlot (2003; BC03), combined with the method from our previous studies (Finkelstein et al. 2007, 2008, 2009).  Briefly, we compute a grid of models, varying the metallicity, star formation history (SFH), stellar population age, and dust extinction (using the dust law of Calzetti et al. 1994).  The metallicities were allowed to be the six discrete values available in BC03, from 0.005 -- 2.5Z\sol, while the SFHs were exponentially decaying, with five values of the characteristic decay time $\tau_{SFH}$ ranging from 10$^{5}$ - 4 $\times$ 10$^{9}$ yr. The age grid was composed of 62 out of the 244 possible ages in BC03, ranging from 1 Myr, up to the age of the Universe at z $\sim$ 0.3 ($\sim$ 10 Gyr).  \lya~and $H\alpha$ emission are included using the number of ionizing photons (output by BC03) and case B recombination.  Various levels of interstellar medium clumpiness (i.e. Neufeld 1991; Hansen \& Oh 2006; Finkelstein et al. 2008, 2009) are allowed by changing the effective dust optical depth for \lya~relative to the continuum. We do this by multiplying the continuum flux by e$^{-\tau}$ and the \lya~flux by e$^{-q\tau}$, where $\tau$ is the optical depth due to dust, and q specifies the level of ISM clumpiness (where q = 0 simulates extremely clumpy, and q = 10 simulates homogeneous).  This effectively decouples the \lya~line strength from the SED fitting, such that the line does not force the best-fit population to be younger than the continuum would indicate.

The model spectra are then redshifted to the {\it GALEX} spectroscopic redshift of a given object (in redshift bins of 0.02), and attenuated by intergalactic medium (IGM) absorption via the Madau (1995) prescription.  The best-fit model is found via $\chi^{2}$ minimization among the five free parameters, using the {\it GALEX} Ly$\alpha$ line fluxes, FUV and NUV fluxes for all objects, along with the optical broadband data discussed above.  The model fitting is independent of the mass, as we fit flux ratios (f$_{x}$/f$_{i}$) to find the best-fit model, and then derive the mass by the weighted mean of the observed optical fluxes to the best-fit model fluxes.  See Finkelstein et al. (2008, 2009) for more extensive details on the model fitting process.  
\begin{figure*}
\epsscale{1.8}
\plottwo{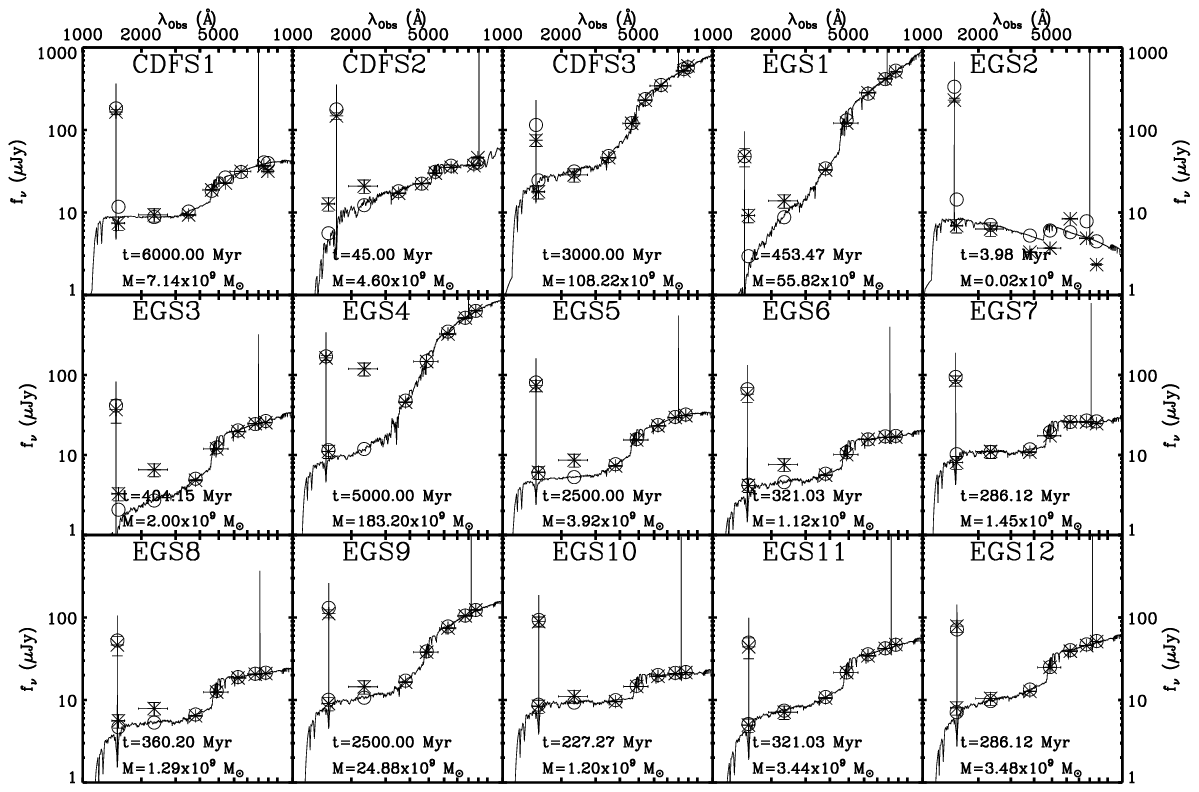}{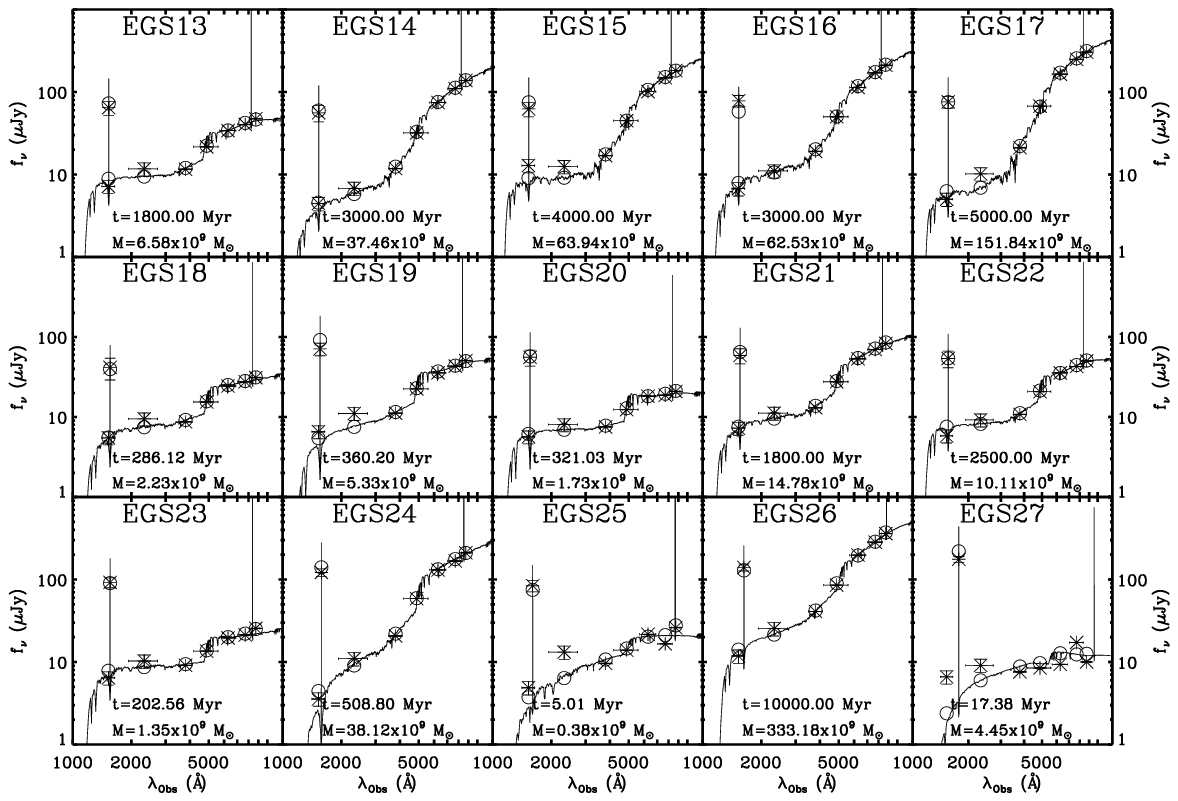}
\vspace{0.5cm}
\caption{Best-fit single population models to the z $\sim$ 0.3 LAEs, with their observed SEDs over-plotted.  The stellar population modeling procedure is discussed in \S 2.1.  The open circles represent the bandpass-averaged fluxes of the models, while the asterisks represent the observed data points.  We fit the models to the GALEX derived \lya~line, NUV and FUV fluxes, as well as optical fluxes from the MUSYC and CFHTLS surveys for the ECDF--S and EGS objects, respectively.  The emission lines shown in the models are \lya~and H$\alpha$.}\label{sed}
\end{figure*}

In order to estimate uncertainties in our results, we followed the method prescribed by Kauffmann et al. (2003), which utilizes Bayesian likelihoods to estimate model-fitting parameter uncertainties.  The most probable value of a parameter is thus the peak of the probability distribution, P $\propto$ e$^{-\chi^{2}}$.  This is simple for the one parameter case, but we have five free parameters.  Thus, to find the collective error in one parameter, we marginalize over all other parameters, in effect collapsing the errors down into one dimension.  This equates to assuming a uniform prior probability distribution in these other parameters.  There is no requirement on the shape of this probability distribution, thus the 1 $\sigma$ error is the range of parameters once the 16\% tails on each end are removed.  Figure 3a shows the age probability distribution of one of our objects.  This method is only useful for those parameters which have a quasi-uniform distribution, thus we are restricted to obtaining parameter errors to the stellar population age, dust extinction and q.  We also report errors on our derived masses, however given that our best-fit masses came from fitting the colors of the objects, we have no mass $\chi^{2}$ distribution.  Rather, in order to ensure that the uncertainty in all of the parameters comes through to the uncertainty in the mass, we ran 10$^{3}$ Monte Carlo simulations, varying the flux by an amount proportional to the flux error (inclusive of the zeropoint error; see \S 3).  The quoted mass error is the range of the middle 68\% of the 10$^{3}$ best-fit masses.

Prior to model fitting, it is important to ensure that the fluxes being used are consistent.  Typically, one would prefer to use the total flux, which can be approximated by the SExtractor MAG\_AUTO extraction.  Both the CFHTLS and MUSYC fluxes used in the paper come from the MAG\_AUTO measurement.  While the GALEX FUV and NUV fluxes have a photometric accuracy of 0.05 and 0.03 mag, respectively (Morrissey et al. 2007), these measurements can be highly uncertain given the broad FWHM of the GALEX beam.  We thus assume a standard error of 0.2 mag for both the GALEX FUV and NUV fluxes, which should result in their being assigned a lesser weight during the fitting process, as befits their more uncertain nature.  In addition, while we expect the Madau IGM prescription to be valid at these redshifts, if the IGM is different at z $\sim$ 0.3, this would affect primarily only the FUV flux, showing a further need to reduce its weight in the fit.  As we mention above, we also use the GALEX measured \lya~line flux measurement in our model fitting, which we believe is well calibrated due to the high throughput of the GALEX prism.  D08 measure a 1 $\sigma$ precision on the line-flux measurement of $\times$ 10$^{-16}$ erg s$^{-1}$ cm$^{-2}$, which we use during our fitting process. 

\section{Results}
Table 1 tabulates and Figure 2 displays the best-fit models for each of our objects.  We gave our objects new identifiers, with CDFS1--3 representing the three LAEs in the ECDF--S, and EGS1-27 the 27 LAEs in the EGS.  While we also fit a model allowing two separate bursts of star formation (one at 10 Gyr ago, and one at any time; see Finkelstein et al. 2009 for details), these fits were ruled out by a large margin in most cases, only providing a marginally better fit to two objects (CDFS3 and EGS1)\footnote[2]{In the limit of no mass in the old population, these models are not equal to the single-population models, as these models are fixed to have a ``burst'' SFH, which we model with $\tau_{SFH}$ = 10$^{5}$ yr.}.  Since we are mainly interested in the average properties of these LAEs, we thus move forward using only the single population models.  

During our initial model fitting, the lowest reduced $\chi^{2}$ ($\chi^{2}_{r}$) was still much greater than one, even though many of the best-fit models appeared to match the observations.  We found that our flux errors were likely underestimated, as these objects are so bright, that their flux errors are rather small.  To correct for this, we added a systematic error equal to the uncertainty in the photometric zeropoint for each band into the $\chi^{2}_{r}$ denominator, which allowed for the objects with the best fits to have $\chi^{2}_{r}$ near to unity (e.g., Papovich et al. 2001).  These errors were 1.2, 0.9, 0.4, 0.5, 0.6 and 1.4\% for the MUSYC U, B, V, R, I and z' bands, respectively, and 2\% for all of the CFHTLS bands.  We attempted to rule out the possibility that any of these objects were AGN by examining X-ray data where available (AEGIS {\it Chandra} data) to see if any of our objects were detected.  We found one object which had a nearby X-ray counterpart; EGS9, with an X-ray source within 0.2\arcs.  However, the full-band X-ray luminosity of this source is L$_{FB}$ = 8.4 $\times$ 10$^{41}$ erg s$^{-1}$, which could indicate a low-luminosity AGN, but it is also consistent with X-ray levels from star-forming galaxies (e.g., Szokoly et al.\ 2004), thus we leave this object in our sample.  It is worth noting that a few other galaxies show possible signs of AGN.  EGS4 exhibits a very steep, red spectral slope, reminiscent of a power-law.  By itself, this could just be indicative of heavy reddening.  However, the inability of the model to fit the {\it GALEX} NUV flux could indicate a possible emission line at that wavelength, possibly HeII $\lambda$1640 or CIII $\lambda$1909, both of which could be present in an AGN.  Thus, EGS4 could be a heavily obscured quasar (e.g., Gregg et al. 2002; Kuraszkiewicz et al. 2009), however we require follow-up optical spectroscopy to know for sure.  There are a few other objects where the NUV flux is difficult to fit, and nebular continuum emission could also be expected to contribute at these wavelengths, although this effect is likely only significant in metal-poor galaxies within a few Myr of a recent starburst (Schaerer 2002).

\subsection{Stellar Population Age and Mass}
From our model fitting, we find an age range of 4 Myr -- 10 Gyr, with a median age of $\sim$ 0.5 Gyr.  Correspondingly, we derive stellar masses from 1.7 $\times$ 10$^{7}$ -- 3.3 $\times$ 10$^{11}$ $M$\sol, with the lowest masses corresponding to the youngest ages, as we would expect.  The best fits of the four youngest objects, EGS2, EGS25, EGS27 and CDFS2 are all of poor quality ($\chi^{2}_{r}$ $\gtrsim$ 30).  Along with the fact that the next youngest object is 200 Myr old, this casts doubt on the results from these four objects.  These four objects also have four of the five highest values of $\chi^{2}_{r}$ (CDFS1 also has $\chi^{2}_{r}$ $>$ 30).  Excluding these four objects from the rest of our analysis, we find age and mass ranges of 203 Myr -- 10 Gyr (median $\sim$ 1.8 Gyr) and 1.1 $\times$ 10$^{9}$ -- 3.3 $\times$ 10$^{11}$ $M$\sol (median $\sim$ 7 $\times$ 10$^{9}$ M\sol), respectively.  This means that the stars in the oldest galaxies in our sample formed as long as 10 Gyr ago, or nearly the age of the Universe at these redshifts.

In Finkelstein et al. (2009) we fit stellar populations to 14 z $\sim$ 4.5 LAEs, and found an age and mass range of 3 -- 500 Myr and 1.6 $\times$ 10$^{8}$ -- 5.0 $\times$ 10$^{10}$ $M$\sol, respectively.  The lower end of our new results is consistent with our previous work, while the upper end represents much more evolved objects, with masses up to 6.7 times greater.  The median stellar mass at z $\sim$ 0.3 is $\sim$ 7.5 times greater than the median mass we measured at z $\sim$ 4.5.  As for age, the median age at z $\sim$ 4.5 was 12 Myr (with the oldest stellar population age of 500 Myr), while at z $\sim$ 0.3, half of the objects are older than 1 Gyr.

The error analysis discussed above allows us to examine the age likelihood distribution of individual objects, by plotting a histogram for each object to see how many simulations resulted in a given age (Figure 3a).  We can thus analyze the sample as a whole by making this plot for each object, and taking the average, which is what we show in Figure 3b.   We find that an average LAE at z $\sim$ 0.3 is highly likely to have an age $>$ 100 Myr, with with many $>$ 1 Gyr.  
\begin{figure*}[th]
\epsscale{1.1}
\plottwo{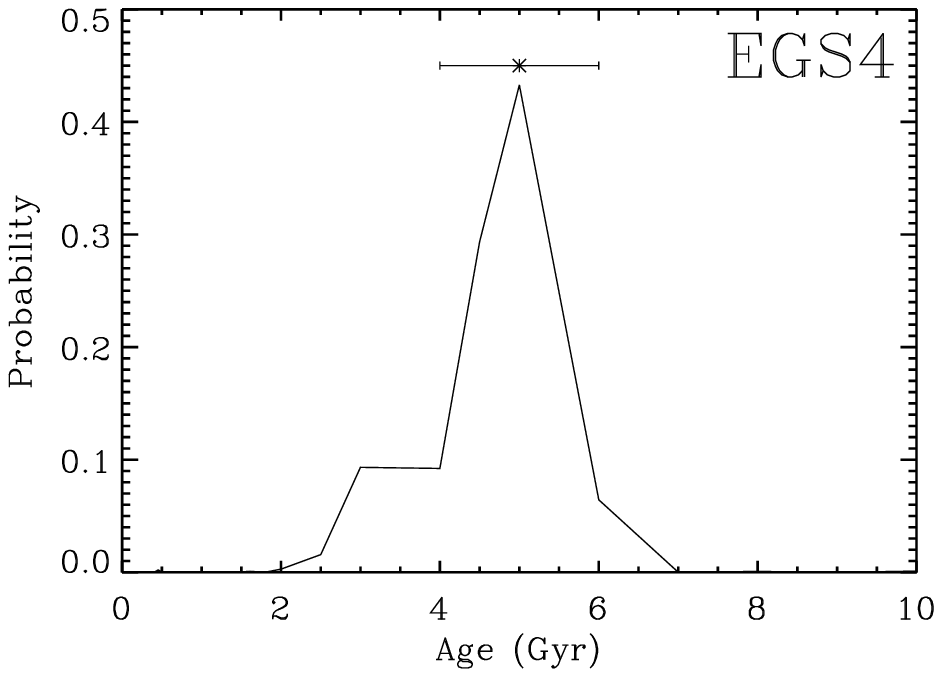}{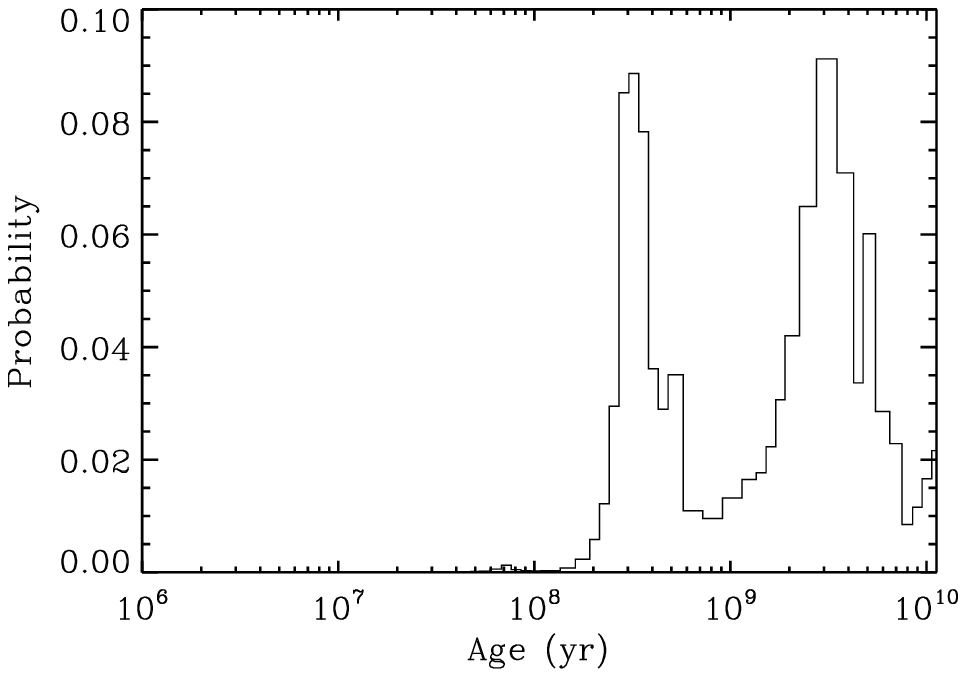}
\caption{(a) The age probability distribution for one of our objects.  The asterisk represents the best-fit model (from minimizing $\chi^{2}$), while the error bars represent the 1 $\sigma$ range from the probability curve.  (b)  The distribution of ages from the probability curves averaged over all 30 objects, showing that LAEs at low redshift are likely to be $>$ 200 Myr old.}\label{agemc}
\end{figure*}

Many other studies at high redshift have derived the physical properties of LAEs via similar methods, although the majority of these studies use stacking analyses to derive average properties of the sample, which can wash out individual results.  In Gawiser et al. (2006a), from a stack of $\sim$ 18 LAEs spectroscopically confirmed to be at z $\sim$ 3.1, they found a typical age of 90 Myr, and a typical mass of 5 $\times$ 10$^{8}$ M\sol.  Also at z $\sim$ 3.1, Nilsson et al. (2007) find a mass of $\sim$ 10$^{9}$ M\sol~from a stacking analysis of 24 LAEs, with an age from a few to several hundred Myr (the age was not well constrained).  Using {\it Hubble Space Telescope} grism spectroscopy to discover 10 LAEs at z $\sim$ 5, Pirzkal et al. (2007) found ages of only a few million years and very low masses, from 10$^{6}$ -- 10$^{8}$ M\sol.  Lastly, Pentericci et al (2009) studied Lyman break galaxies (LBGs) with \lya~emission (which we can also consider to be LAEs, although the LBG selection criteria preferentially selects brighter, and thus more massive galaxies), and found stellar masses from $\sim$ 5 $\times$ 10$^{8}$ -- 5 $\sim$ 10$^{10}$ M\sol.  Additionally, they found that while the majority of their galaxies were best-fit by young populations, a small fraction required a significantly older component.  Overall, the literature is consistent with what we have found at z $\sim$ 4.5, in that high-redshift LAEs appear to be mainly young and low-mass, although some small fraction do appear to be more involved.  The evolution of this fraction with redshift is unknown, and is a top priority question for future studies.

\subsection{Metallicity, Dust and ISM Clumpiness}
The model grid consisted of six allowed values of metallicity:  0.005, 0.02, 0.2, 0.4, 1.0 and 2.5 $Z$\sol.  Only two objects (CDFS1 and EGS26) out of 26 was best fit with the lowest value of metallicity, with five more being best fit with Z = 0.02 $Z$\sol.  The majority of objects (19) were best fit with Z $\geq$ 0.2 $Z$\sol.  In F09, we found that at z $\sim$ 4.5 50\% of our objects were fit with metallicities of 0.02 $Z$\sol~or less, thus showing possible evidence of metallicity evolution with redshift.  This evolution does not show up in the dust content, as the mean extinction at z $\sim$ 0.3 is \dust~$\simeq$ 1.1 mag, while at z $\sim$ 4.5 it was \dust~$\simeq$ 2.5 mag.  While the ISM clumpiness (q) spans a similar range at low redshift as at z $\sim$ 4.5, at z $\sim$ 4.5, over 50\% of the objects had q $\leq$ 1.  At z $\sim$ 0.3, only nine objects have dust enhancement of the \lya~EW occurring with q $<$ 1, but dust is not extinguishing the EWs of these objects much either, as 22/26 objects have q $\leq$ 1.5.  Histograms comparing the age, dust, metallicity and clumpiness differences between these two redshifts are given in Figure 4.  We computed the Kolmogorov-Smirnov (KS) statistic, using the IDL task kstwo, to explore these histograms further.  Using age as an example, the KS statistic computes the probability that the array of age values at z $\sim$ 4.5 and z $\sim$ 0.3 are drawn from the same distribution.  This returned a probability of 7.5 $\times$ 10$^{-7}$, thus there is a very low chance that the ages at the two redshifts arise from the same distribution.  Similarly, we find low KS probabilities of 0.003, 0.102 and 0.007 for A$_{1200}$, Z and q respectively, with the slightly higher probability for Z possibly due to the very coarse grid of possible metallicities.  It is comforting that the shape of Figure 3b, and the first panel of Figure 4 are similar, as it shows that our ages accurately represent the most probable results.

While Gawiser et al. (2006a) and others (e.g., Lai et al. 2008) derive A$_{V}$ consistent with zero in their studies, many recent studies do find dust in their LAEs\footnote[3]{With the Calzetti et al. (1994) dust law, A$_{1200}$ $\sim$ 4 $\times$ A$_{V}$.}.  Even though their objects were extremely young, Pirzkal et al. (2007) found dust in up to half of their LAEs, with A$_{V}$ as high as 0.6, while Nilsson et al. (2007) found A$_{V}$ $\sim$ 0.3.  Pentericci et al. (2009) also found dust in over half of their objects.  These results are also consistent with our results at z $\sim$ 4.5, although the uncertainties in the dust measurements at high redshift are large (given that most of these studies only loosely constrain the 4000 \AA~break), thus any evolution in the dust content in LAEs are uncertain.
\begin{figure*}
\epsscale{1.0}
\plotone{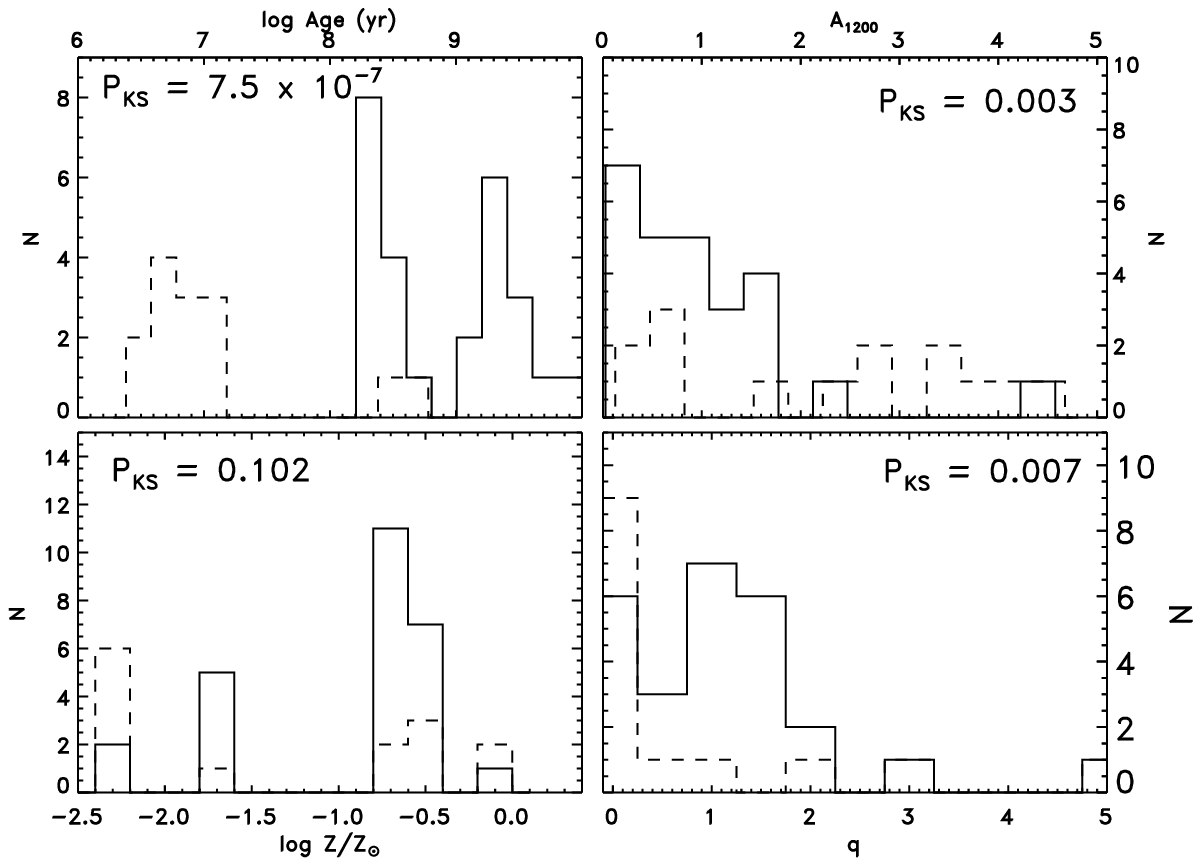}
\vspace{0.7cm}
\caption{Histograms showing the age, dust, metallicity and ISM clumpiness distributions for this z $\sim$ 0.3 sample (solid lines) and our z $\sim$ 4.5 sample (dashed lines).  Strong differences are seen in all four physical properties, strongly implying that low-redshift LAEs are different than their high-redshift counterparts.  We list the results of KS tests done on each of the following distributions, showing the probability that the results at the two redshifts are drawn from the same distribution.}\label{agemc}
\end{figure*}

These differences in metallicity and ISM properties at different redshifts may give us insight into the evolution of these objects.  We would expect the average stellar population to become more metal enriched with time, and we believe that is what we are seeing with the metallicity differences between z $\sim$ 4.5 and 0.3.  The fact that objects at z $\sim$ 4.5 have more dust extinction on average is interesting, although it may just be a selection effect.  The z $\sim$ 0.3 LAEs were selected via spectroscopy, thus they were more sensitive to low EW objects.  However, at z $\sim$ 4.5, we required a significantly strong narrowband excess in order to select an object as an LAE, which biased us towards higher values of the \lya~EWs.  As we showed in F09, a large portion of our sample had high EWs possibly due to clumpy dust enhancement, thus we may have been biased towards dusty objects as well.  According to the best-fit models, only 36\% of our z $\sim$ 4.5 LAEs had EW $<$ 120 \AA, whereas as z $\sim$ 0.3, over 55\% of our LAEs have EWs $<$ 120 \AA.  In fact, for z $\sim$ 0.3, the measured EWs from D08 may be more accurate as they come from spectroscopy.  Using these numbers, all 30 z $\sim$ 0.3 LAEs have EWs $<$ 120 \AA, with a median value of 31 \AA, and a minimum of 11.8 \AA.  The change in EW with redshift could also be due to the apparent evolution of these objects, since EW is a ratio between the line flux and continuum flux.  If the amount of Ly$\alpha$ photons being produced is similar, but the continuum is brighter, the EW will drop.  At high redshift, LAEs are known to be relatively small, with r$_{e}$ $\leq$ 1 kpc (Pirzkal et al. 2007), thus it is likely that these objects are producing their \lya~photons in HII regions in much smaller disks than at low redshift.  Even if low-redshift LAEs had a similar number of HII regions, they would be spread over a much larger disk, diluting the EW as the larger disk is producing more continuum light.

One may then ask, why do we not also see the high EW, dust enhanced LAEs at low redshift?  There are a few, as discussed in the above paragraphs, but the numbers are significantly smaller than those at high redshift.  This may indicate an evolution in the dust geometry in LAEs, and/or galaxies in general.  Over time, a galaxy will accumulate more and more dust as stars continue to grow old and die.  Isolated regions may become clumpy due to turbulence in H II regions among other things.  In the high-redshift LAEs, which may be intrinsically smaller, these local events may be enough to change the ISM on a galaxy wide scale, allowing for the large range in q we found.  However, at z $\sim$ 0.3, these objects may be larger in size, thus it might be harder for a whole galaxy to have an entirely clumpy or entirely homogeneous ISM, which may be responsible for the fact that most z $\sim$ 0.3 LAEs have q near 1 -- 1.5.  In addition, external events, such as galaxy mergers, can severely disrupt the ISM.  Ryan et al. (2008) show that the major merger fraction peaks at z $\sim$ 1.3.  If the time since their last merger is large, this will also allow low-redshift LAEs to have more settled ISMs.

Although not much work has been done on LAEs at low redshift, a recent study has been published studying the role of dust in local \lya~emitting galaxies (Atek et al. 2008).  Specifically, their plots of \lya~EW vs. dust extinction allow an opportunity to check if clumpy dust is affecting the EW, and by how much.  In Figure 3 of their paper, they show this plot for Haro 11.  The emission component of this plot shows significant scatter at low E(B-V), however, it approaches a constant value of $\sim$ 20 \AA~from E(B - V) = 0.5 -- 1.0.  In order for the \lya~EW to stay constant with increasing dust content, q must equal 1, which is the most likely value we found for q at z $\sim$ 0.3.  This same effect is seen in their plots of IRAS 08339+6517, and also in NGC 6090, albeit less significantly.  We believe that this confirms our prediction that the ISM of lower-redshift LAEs are neither extremely clumpy nor entirely homogeneous.

\subsection{Stellar Evolution Uncertainties}
While the BC03 models are very widely used across extragalactic astrophysics, recent insights into the contributions of post main-sequence evolution to the overall SED of a galaxy has called these models into question.  The BC03 models do include all phases of stellar evolution from the zero-age main-sequence until the stellar remnant stage, including the thermally-pulsating asymptotic giant branch (TP--AGB) phase.  However, Maraston et al. (2005, 2006) have recently shown that the BC03 models underestimate the level at which TP-AGB stars can dominate the rest-frame near-IR ($\lambda \gtrsim$ 1 $\mu$m), causing overestimates of the derived ages and masses.  

Even prior to our model fitting we would expect these z $\sim$ 0.3 LAEs to be older, possibly in the range where TP--AGB stars can dominate their NIR ($\sim$ 0.2 - 2 Gyr). However, our reddest band (z') corresponds to rest-frame $\sim$ 7000 \AA, thus we would not expect a different treatment of TP--AGB stars to significantly alter our results.  Charlot \& Bruzual (2007 in prep) have devised new models, including the Maraston et al. treatment of TP--AGB stars, which will be released in the near future.  Future studies at redder wavelengths will need to use these, or similar, models to more accurately fit their objects.

\section{Conclusions}

We have performed stellar population model fitting on a sample of 30 GALEX discovered z $\sim$ 0.3 \lya~emitting galaxies.  We derived the stellar population ages and masses, as well as  metallicities, dust extinction, and ISM geometries.  The age and mass results imply that low-redshift LAEs are both older and more massive than their high-redshift counterparts, with the youngest low-redshift LAE having a best-fit age of $\sim$ 200 Myr, much older than the average age at high redshift.  We also found that the metallicity in LAEs is higher at low redshift, showing evolution of the mean metallicity of the stars inside these galaxies, as would be expected over a period of $\sim$ 9 Gyr of cosmic time.

Perhaps most interesting, we found that the ISM at low redshift appears to be very stable, in a quasi-clumpy state.  At high redshift, we found that the ISM in LAEs ran the whole range from extremely clumpy to very homogeneous.  At z $\sim$ 0.3, we found that while a few galaxies were fairly clumpy, most resided in the regime of q $\sim$ 1, where dust will not affect the \lya~EW.  This result is confirmed by the study of local LAE analogs by Atek et al (2008).  Our results show that LAEs at z $\sim$ 0.3 show significant differences from their high-redshift cousins, although further study of a larger sample will help confirm this.  Although low-redshift LAE studies are difficult as they need to be done from space, help should be on the way with the near-UV sensitive Wide Field Camera 3 (WFC3) due to be installed on the {\it Hubble Space Telescope}.

\acknowledgements
We thank the anonymous referee for his/her suggestions, which greatly improved this work.  We thank Rogier Windhorst, Evan Scannapieco, Jean-Michel Deharveng and Casey Papovich for helpful discussions on these topics.  This work was supported by the Arizona State University School of Earth and Space Exploration.

\end{document}